\documentstyle[aps,prl,eqsecnum,preprint,cite,epsf]{revtex}
\tighten

\begin{document}
\title{
Exchange Monte Carlo Method and  \\
Application to Spin Glass Simulations
}
\author{Koji {\sc  Hukushima}}
\address{Institute of Physics, University of Tsukuba, Tsukuba 305, Japan}
\author{Koji {\sc Nemoto}}
\address{Department of Physics, Faculty of Science, Hokkaido University,
Sapporo 060, Japan}
\maketitle
\begin{abstract}
We propose an efficient Monte Carlo algorithm
        for simulating a ``hardly-relaxing" system,
        in which many replicas with different temperatures are simultaneously
        simulated and a virtual process exchanging configurations of these
        replica is introduced.
This exchange process is expected to let the system at low temperatures escape
        from a local minimum.
By using this algorithm the three-dimensional $\pm J$ Ising spin glass model
        is studied.
The ergodicity time in this method is found much smaller than that of
        the multi-canonical method.
In particular the time correlation function almost follows
        an exponential decay whose relaxation time is comparable
        to the ergodicity time at low temperatures.
It suggests that the system relaxes very rapidly through the exchange process
        even in the low temperature phase.
\end{abstract}
\pacs{PACS: }

\section{Introduction}
The low temperature phase of spin glasses (SG)
         and other complex systems generally
        have numerous local minima which are separated to each other
        by energy barriers.
In studying such systems,
        we have to take account of each configuration for these local minima
        and the fluctuation around it.
The characteristic time in which the system escapes from a local minimum,
        however, increases rapidly as temperature decreases.
This situation causes ``hardly-relaxing" problem in using
        conventional Monte Carlo(MC) simulations based on a local updating.

So far, various improvements have been (and is being)
        made on MC algorithms to overcome this problem.
There are mainly two categories in such algorithms, i.e.,
        the cluster algorithm  and the extended ensemble method.
The Swendsen-Wang\cite{Swendsen} and the Wolff\cite{Wolff}
 algorithms are in the first category in which
        non-local updating is employed to let the system easily change over
        energy barriers.
The second one is based on the idea that an extension of the canonical
        ensemble can be introduced so as to perform MC sampling efficiently
        at low temperatures.
The multi-canonical method\cite{Berg1,Berg2}
        and the simulated tempering\cite{Marinari} are two such algorithms.
Some applications of the SG system have been attempted
            for the multi-canonical method\cite{Berg3,Berg4,Berg5}
        and for the simulated tempering\cite{Kerler,Tempering2}.
In this work  we propose another efficient algorithm
        belonging to the second category.
It can be regarded as an parallelized version of the simulated tempering,
        although the extended ensemble used is very different.

Our method treats a compound system which consists of many
        replicas of the system concerned.
The temperature attributed to each replica is distributed in a range including
        both high and low temperature phases.
To this system we perform a standard MC simulation in the following
        way: (1) Each replica is simulated {\it simultaneously} and
        {\it independently} as canonical ensemble for a few MC steps(MCS).
(2) Exchange of the configurations for a pair of replicas is tried
        by referring to the energy cost of the whole system.
In effect, state traverses over the replicas, due to the exchange process,
        being cooled down and warmed up,
        and thus replica at low temperatures can escape from a local minimum
        very easily.
Since the present method never violates the detailed balance
        in the canonical sense,
        each replica is guaranteed to be equilibrated to the canonical
        distribution with its own temperature.
An advantage of our method to the simulated tempering is that we need not
        to estimate weighting factor (or free energy) to equidistribute the
        probability of visiting temperatures.

We then apply the exchange MC method to a three-dimensional $\pm J$ Ising spin
       glass(SG) model and show that it works well even in the SG phase.
To characterize the time scale of relaxation in our method,
        we observe the ergodicity time, which is defined as
        the average MCS per one travel over the whole temperature range,
        and a modified autocorrelation function.
It is found  that the autocorrelation function follows the exponential decay
        and the relaxation time at low temperatures takes
        a moderate value which is comparable with the ergodicity time.
The order parameter distribution $P(q)$ is an even function
        in the whole temperature range,
        which provide us an evidence for reaching the equilibrium.

This paper is organized as follows:
In Sec. 2 we describe the exchange MC algorithm in detail.
In Sec. 3 we show how to determine the temperature mesh
        in order that the exchange process occurs properly.
In Sec. 4 we report the relaxational behavior of our method
        by applying it to the SG model.
The last section is devoted to discussion and summary.

\section{Exchange MC method}
In our method we treat a compound system which consists of non-interacting
        $M$ replicas of the system concerned.
The $m$-th replica, described by a common Hamiltonian ${\cal H}(X)$,
        is associated with inverse
        temperature $\beta_m$, i.e., each replica is in contact with
        its own heat bath having different temperature (for convenience we take
        $\beta_m < \beta_{m+1}$).
A state of this extended ensemble is specified by
        $\{X\}=\{X_1,X_2,\cdots,X_M\}$, and
        the partition function is given as
\begin{equation}
        {\cal Z} = {\rm  Tr}_{\{X\}}\exp(-\sum_{m=1}^{M}\beta_m{\cal H}(X_m))
        = \prod_{m=1}^{M}Z(\beta_m),
\end{equation}
where $Z(\beta)$ is the one for the original system.
For a set of temperature $\{\beta\}$ given,
 the probability distribution of finding $\{X\}$ becomes
\begin{equation}
        P(\{ X,\beta\}) = \prod_m^MP_{\rm eq}(X_m,\beta_m),
\end{equation}
        where
\begin{equation}
        P_{\rm eq}(X,\beta)=Z^{-1}(\beta)\exp(-\beta{\cal H}(X))\label{eqn:Peq}
\end{equation}

In constructing a Markov process for exchange MC we introduce a transition
        matrix $W(X,\beta_m|X',\beta_n)$ which is a probability of exchanging
        configurations of the $n$-th and $m$-th replicas.
In order that the system  remains  at equilibrium, it is sufficient to
        impose the detailed  balance condition on the transition matrix:
\begin{eqnarray}
&&P(\cdots;X,\beta_m;\cdots;X',\beta_n;\cdots)W(X,\beta_m|X',\beta_n)\nonumber\\
&&\qquad= P(\cdots;X',\beta_m\cdots;X,\beta_n;\cdots)W(X',\beta_m|X,\beta_n).
\end{eqnarray}
{}From eq.(\ref{eqn:Peq}) we obtain
\begin{equation}
 \frac{W(X,\beta_m | X',\beta_n)}{W(X',\beta_m | X,\beta_n)}
 =
 \exp (-\Delta),
\end{equation}
where
\begin{equation}
        \Delta=(\beta_n-\beta_m)({\cal H}(X)-{\cal H}(X')).
\end{equation}
Therefore the replica-exchange part of transition probability can be
        expressed as
\begin{equation}
W(X,\beta_m | X',\beta_n)  = \left\{
    \begin{array}{rl}
        1,&\quad\mbox{for $\Delta<0$} \\
        \exp(-\Delta),&\quad\mbox{for $\Delta>0$}
    \end{array}\right.
\end{equation}
if one adopts the Metropolis method.

For the actual MC procedure, the following two steps are performed
        alternately:
\begin{enumerate}
\item Each replica is simulated {\it simultaneously} and {\it independently}
        as canonical ensemble for a few MCS by using a standard  MC method.
\item Exchange of two configurations $X_m$ and $X_{m+1}$,
        is tried and accepted with the probability
        $W(X_m,\beta_m | X_{m+1},\beta_{m+1})$.
\end{enumerate}
Here we restrict the replica-exchange to the case $n=m+1$
        because the acceptance ratio of the exchange trial
        decreases exponentially with the difference
        $\beta_m-\beta_n$.

The canonical expectation value of a physical quantity $A$ is
        measured in usual way as follows:
\begin{equation}
        \langle A\rangle_{\beta_m} =
        \frac{1}{N_{\rm mcs}}\sum_{t=1}^{N_{\rm mcs}}
        A({X}_m(t)).
\end{equation}
Another expression can be obtained when
the exchange procedure mentioned above is regarded as for temperature,
i.e., temperatures, instead of configurations, of a pair of replicas
are to be exchanged.
Then the above quantity is expressed as
\begin{equation}
        \langle A\rangle_\beta =
        \frac{1}{N_{\rm mcs}}\sum_{t=1}^{N_{\rm mcs}}
        \sum_{m=1}^{M}A(\tilde{X}_m(t))\delta_{\beta,\beta_m(t)},
\end{equation}
where we introduce the time-dependent inverse temperature $\beta_m(t)$ and
the configuration $\tilde{X}_m$ in this temperature-exchange scheme.
Note that both schemes are completely equivalent to one another.
One can choose either of the two schemes in actual implementation of
the present method.

\section{determination of temperature}
The parameters we have to determine are only the set of (inverse)
        temperatures $\{\beta_m\}$.
The highest temperature should be set in the high temperature phase
        where relaxation (correlation) time is expected to be very short
        and there exists only one minimum in the free energy space,
        otherwise  the system would not completely forget where it was trapped
        before even if it visits to the highest temperature.
On the other hand, the lowest temperature is somewhere in the low temperature
        phase whose properties we are interested in.
In this sense the temperature range is considered to be given.
Then we estimate the number of temperatures required in the range as follows.
In order that each replica wanders over the whole temperature region
        the  acceptance probabilities of
        the exchange process for  every pair of replicas
        at different temperatures
        have to be of order of one and nearly constant.
The logarithm of the probability ${\rm e}^{-\Delta}$ of exchanging $\beta_n$
        and $\beta_{n+1}=\beta_n+\delta$ is to order $\delta^2$
\begin{equation}
        \Delta=\delta({\cal H}(X_{n+1})-{\cal H}(X_n))
        \sim\delta^2\frac{\rm d}{\rm d\beta}E,
\end{equation}
where the instantaneous value of the energy ${\cal H}$ is approximated by the
        thermal expectation value $E$.
Since the energy $E$ is an extensive variable,
        $\delta$ should be of order of $\frac{1}{\sqrt{N}}$ to satisfy
        the condition that $\Delta\sim O(1)$.
In other words, the number of temperatures in the range we have to simulate is
        of the order of $\sqrt{N}$.
In the case where a second-order phase transition exists, the number is
        modified as $\sqrt{N^{1+\frac{\alpha}{d\nu}}}$,
         where $\alpha$ and $\nu$ are the exponent of the specific heat
        and the correlation length.
This means that we need more replicas if  the specific heat diverges
        at the transition temperature.
In spin glass and other glassy systems, the specific heat has usually
        no singularity at the transition temperature.

A set of temperatures $\{\beta_m\}$ can be obtained by the iteration
        procedure which is first introduced to
        the simulated tempering.\cite{Kerler}
For given $\{\beta_m\}$, the acceptance ratios $\{p_m\}$ are
        evaluated by simulating an appropriate MCS.
Then a new set $\{\beta'_m\}$ is
        constructed by using the old set $\{\beta_m\}$:
\begin{eqnarray}
        \beta'_1 & = & \beta_1, \nonumber \\
        \beta'_m & = & \beta'_{m-1} +
        (\beta_m-\beta_{m-1})\frac{p_m}{c},  \hfill (m=1,2,\cdots,M), \nonumber
\\
        c&=&\frac{1}{M-1}\sum_{m=1}^{M}p_m.
        \label{eqn:iteration}
\end{eqnarray}
In each iteration the MCS required to estimate $p_m$ are found not
        so long.
We will show a concrete examination in the next section.

In closing this section, we propose some necessary conditions to check the
        efficiency and the equilibration.

(i) The exchange happens with a non-negligible probability for all
        adjacent pairs of replicas.

(ii) Each replica moves around  the whole temperature range in suitable MCS.

(iii) In moving the temperature space the system  forgets where
 it was trapped.\\
We can obtain the temperatures $\{\beta_m\}$ satisfying the condition (i)
        by the iteration procedure mentioned above.
The ergodicity time defined in the following section gives a criterion
         for the condition (ii).
Unfortunately, even if the condition (i) and (ii) are satisfied,
 it is possible that the condition (iii) is violated.
One trivial  possibility for this is that the largest temperature is
        not high enough.
If some of these conditions are broken down, we may  not be able to improve
        the present method due to the absence
        of any other controlling parameters.

\section{The model and simulations}
We consider the three dimensional $\pm J$ Ising SG model
        on the simple cubic lattice.
The real-replica Hamiltonian is defined as
\begin{equation}
        {\cal H}(\{\sigma,\tau\})
        = -\sum_{ij}J_{ij}(\sigma_i\sigma_j+\tau_i\tau_j),
        \label{eqn:model}
\end{equation}
        where $\{\sigma\}$ and $\{\tau\}$ take the values $\pm 1$.
The interactions $\{ J_{ij}\}$ are quenched random variables
        taking $\pm 1$ with equal probability.
To be more accurate, $\{ J_{ij}\}$ are distributed so that
        the number of the ferromagnetic bonds, $J_{ij}=+1$,
        is exactly a half of the total bonds.
Then the Edwards-Anderson SG order parameter\cite{EA}
        is computed as the overlap between the two copies,
\begin{equation}
        q = \frac{1}{N}[\sum_i\langle \sigma_i\tau_i\rangle],
        \label{eqn:EA}
\end{equation}
        where $\langle\cdots\rangle$ and $[\cdots]$ denote
        thermal average for the Hamiltonian (\ref{eqn:model})
        and random average, respectively.
We have simulated on lattices of linear size $L=$ 6, 8, 12, and 16 with
    periodic boundary condition.
For local updating we have adopted the conventional sublattice-flipping
        with heat bath method.
For the exchange process
the replica pairs $(\beta_m,\beta_{m+1})$ are divided into two subgroups,
        $i.e.,$ odd-$m$ and even-$m$ groups.
In actual updating the exchange trial is performed for one of
        these subgroups after each local updating.
It means that our one MC step consists of one local updating MC step
        and a half exchange-trial per replica-pair.

\subsection{temperature setting and the ergodicity time }
The actual local updating have been performed with the multi-spin coding
        technique\cite{Zorn,KikuchiOkabe,ItoKanada}
        which simulates 32 different physical systems at once.
This fixes the total number of temperatures to be $M=32$
        in all lattice sizes.
To determine the temperatures $\{\beta_m\}$
        a typical sample of size $L=12$ has been used.
The initial condition we used is
\begin{equation}
\beta_m = \beta_1 + (\beta_M-\beta_1)\frac{m-1}{M-1},
\end{equation}
with $\beta_1=0.4$ and $\beta_M=2$.
We then performed 400 MC steps of simulation for each iteration of
        eq.(\ref{eqn:iteration})
to evaluate the acceptance probabilities $\{p_m\}$,
and found that
the temperatures converge after several iterations.
Finally we smoothed $\{\beta_m\}$ by spline interpolation and
        obtained the temperature range $(0.86\le T/J\le 2.39)$
         which includes both the high and low  temperature phases.
These temperatures so obtained was used for all samples and sizes.
In fig. \ref{fig:accept} we show the acceptance ratios for various sizes.
These values are found to be of order of one and do not depend on bond
        realization so sensitively.

In order to confirm that the obtained temperature set satisfies the condition
        (ii) mentioned in the previous section, we investigated the ergodicity
                time $\tau_{\rm E}$
       which is defined as the average MC step for a specific replica to move
                from the lowest to the highest temperature.
The dependence of our observed ergodicity time on the lattice size is
       shown in Fig. \ref{fig:tunnel},
       with that of the multi-canonical method by Berg et al.\cite{Berg5}
If the total number of  temperature points we simulate is fixed as in the
        present analysis,
    the acceptance ratio, e$^{-\Delta}$,  of exchange process between replicas
    reduces exponentially as a function of system size $L$
    and the  $\tau_{\rm E}$ grows rapidly with increasing $L$.
As shown in Fig. \ref{fig:tunnel}, however, it is clear that
         our ergodicity time is much shorter than
         the multi-canonical ergodicity time in a  lattice size suitable
         for our actual simulation.

\subsection{relaxation}
Because each replica wanders over the temperature space,
        the (equilibrium) time correlation function can not be argued in the
        ordinary sense.
Here to investigate relaxation dynamics of the present method,
        we study an autocorrelation function
        in the temperature-exchange scheme,
\begin{equation}
        q(t,\beta_m) = \frac{1}{N}\sum_i\langle
        \tilde{\sigma}_i(0)\tilde{\sigma}_i(t)\rangle_{\rm path}^{\rm (m)},
\end{equation}
where $\langle\cdots\rangle_{\rm path}^{\rm (m)}$ means an average along
trajectories
        in temperature space evolving
        from the initial state $\{\tilde{\sigma}_i(0)\}$
        with temperature $\beta_m$.
The function $q(t,\beta)$ is expected to include the slowest relaxation mode
        of the present method.
In Fig. \ref{fig:qoft} we show $q(t,\beta)$ at various temperatures
        for $L=12$.
The data are averaged over 10 samples.
As indicated by the straight lines,
  the function $q(t,\beta)$ follows  nearly exponential
        decay for large time $t$.
We evaluate the largest relaxation time from least square fit to
        a single exponential
\begin{equation}
        \label{eqn:fit}
        q(t,\beta) \sim q_0\exp(-t/\tau)
        \ \ \ \ t \gg 1.
\end{equation}
As shown in Fig. \ref{fig:fit}(a) the relaxation time $\tau$ has a
     crossover temperature $T_{\rm cr}\sim 1.5$, above which
     $\tau$ behaves like the relaxation time obtained by a
     conventional local spin flip dynamics.
Below $T_{\rm cr}$, on the other hand, $\tau$ seems to almost become
         saturated and  to be comparable with $\tau_{E}$,
         and the relaxation amplitude
     $q_0$ grows rapidly (see Fig. \ref{fig:fit}(b)).
It suggests that the largest relaxation mode of the exchange dynamics
     dominates the relaxation at low temperatures.
In any case we can see that the system relaxes within a reasonable MCS and
     thus the condition (iii) is satisfied.

\subsection{order parameter distribution}
\begin{figure}
\caption{Distribution function $P_J(q)$ of the overlap for
        a sample with $L=16$ at $T=0.924$. The solid
        and dashed line represent $P(q)$ by using the exchange MC
        and by a conventional MC method, respectively.
        The same MCS are used in two methods. }
\label{fig:demo}
\end{figure}
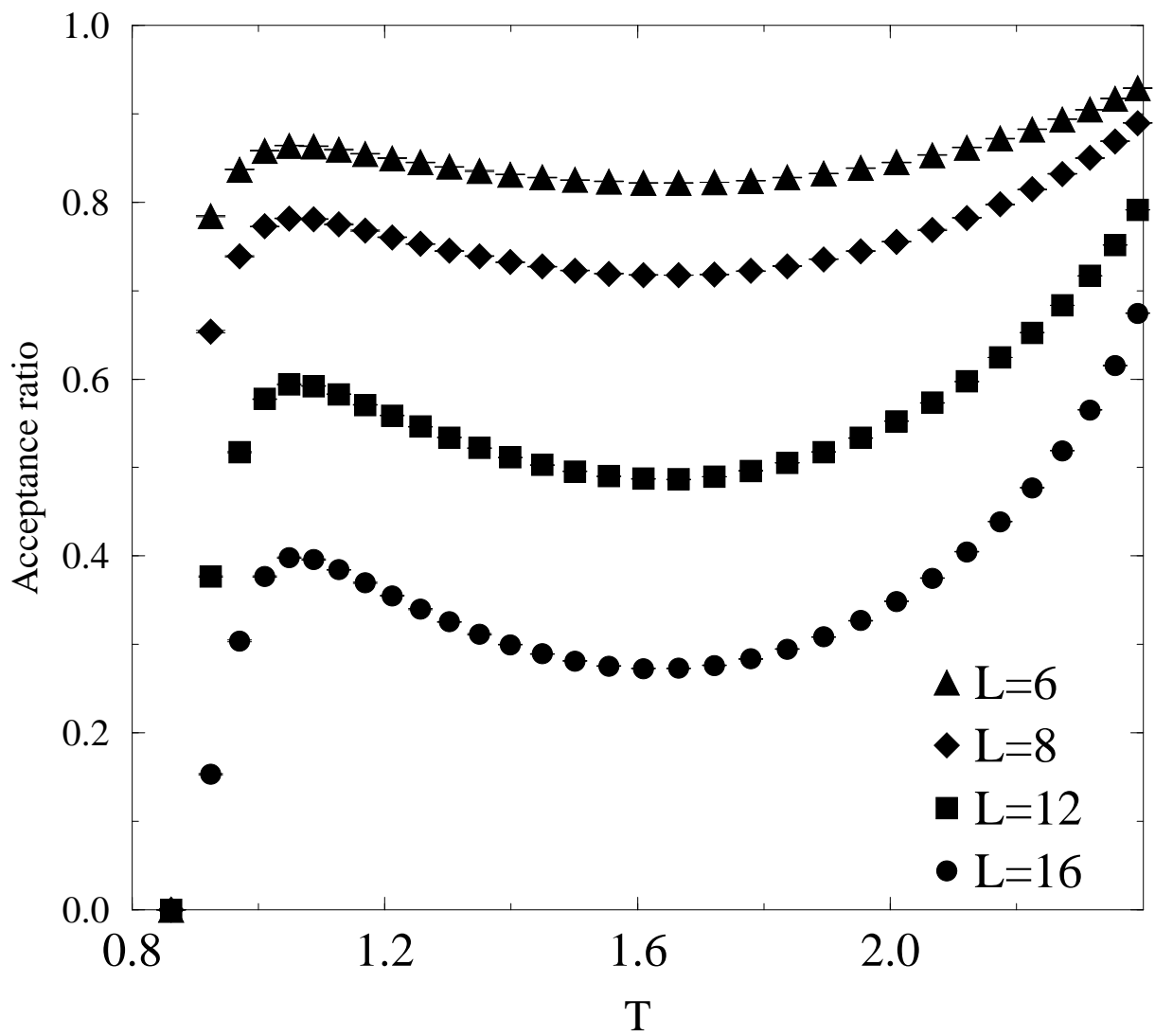
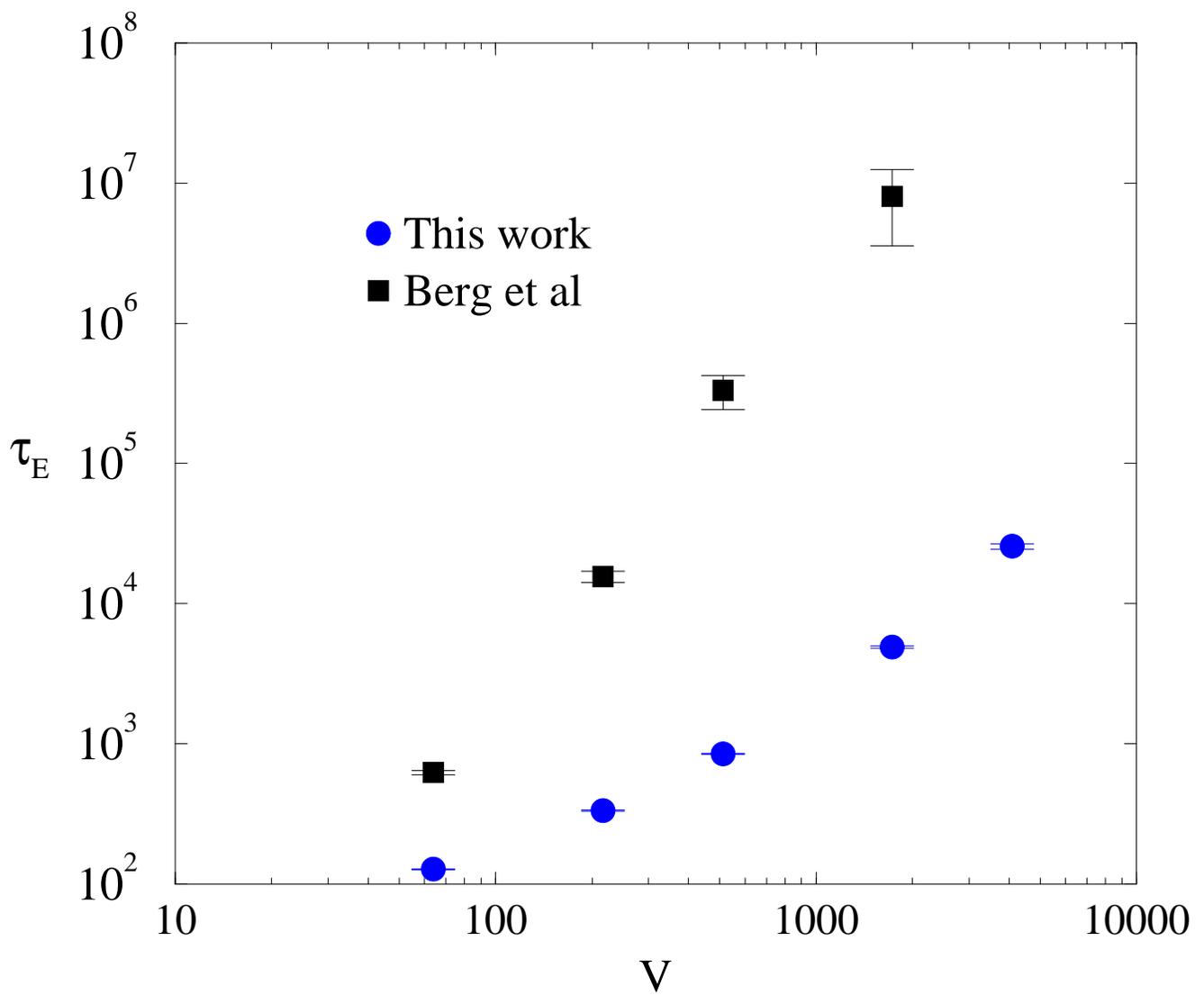
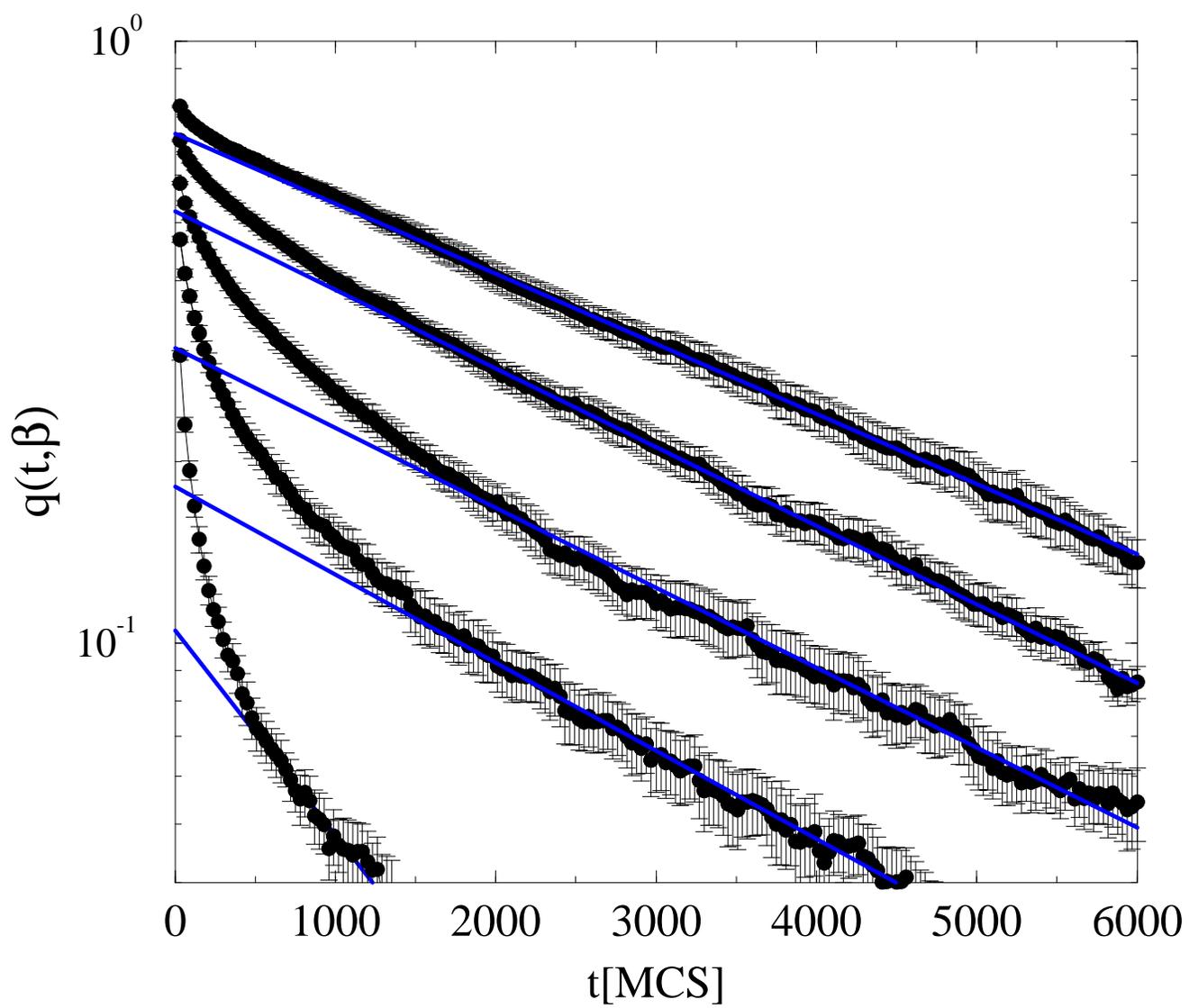
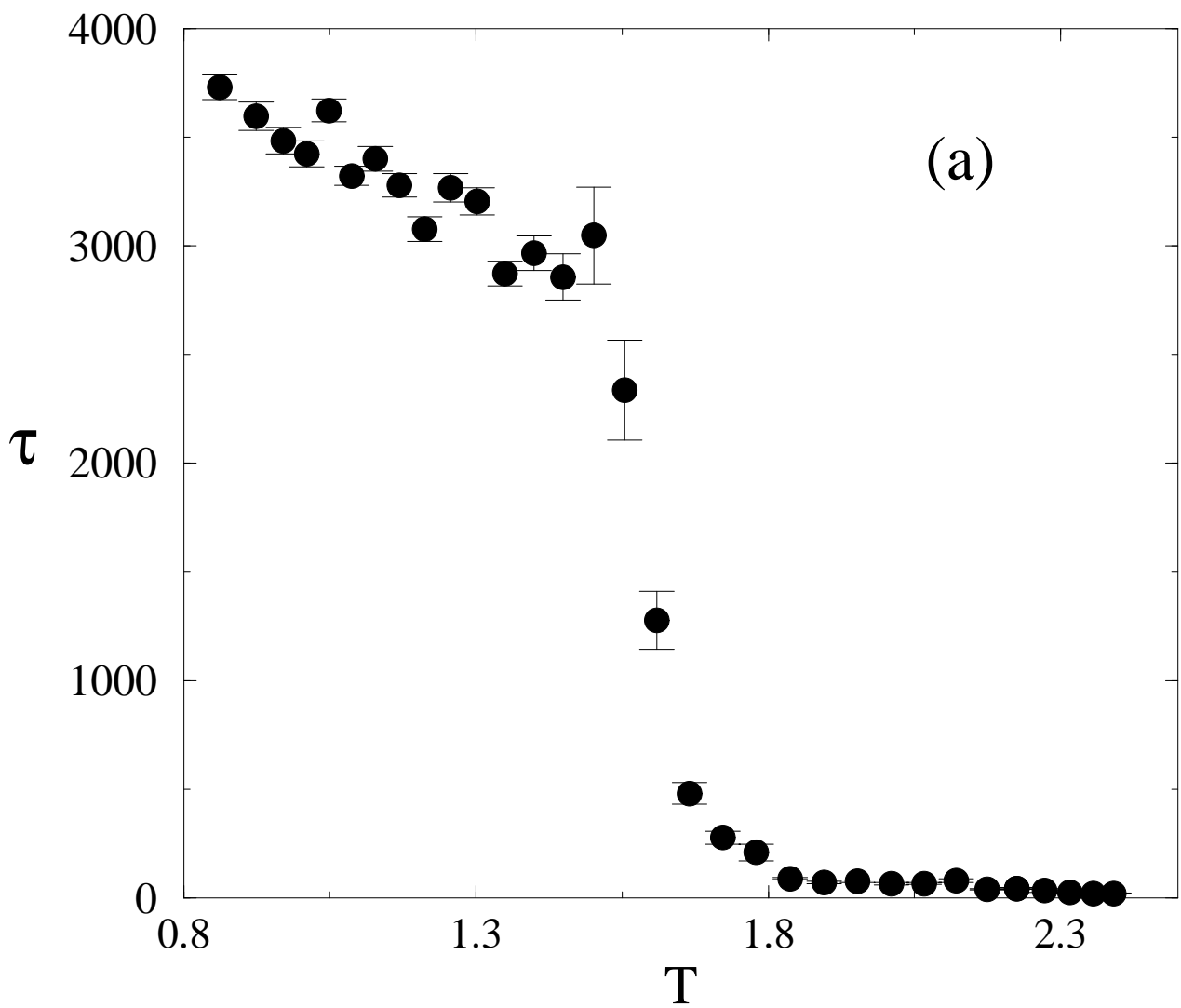
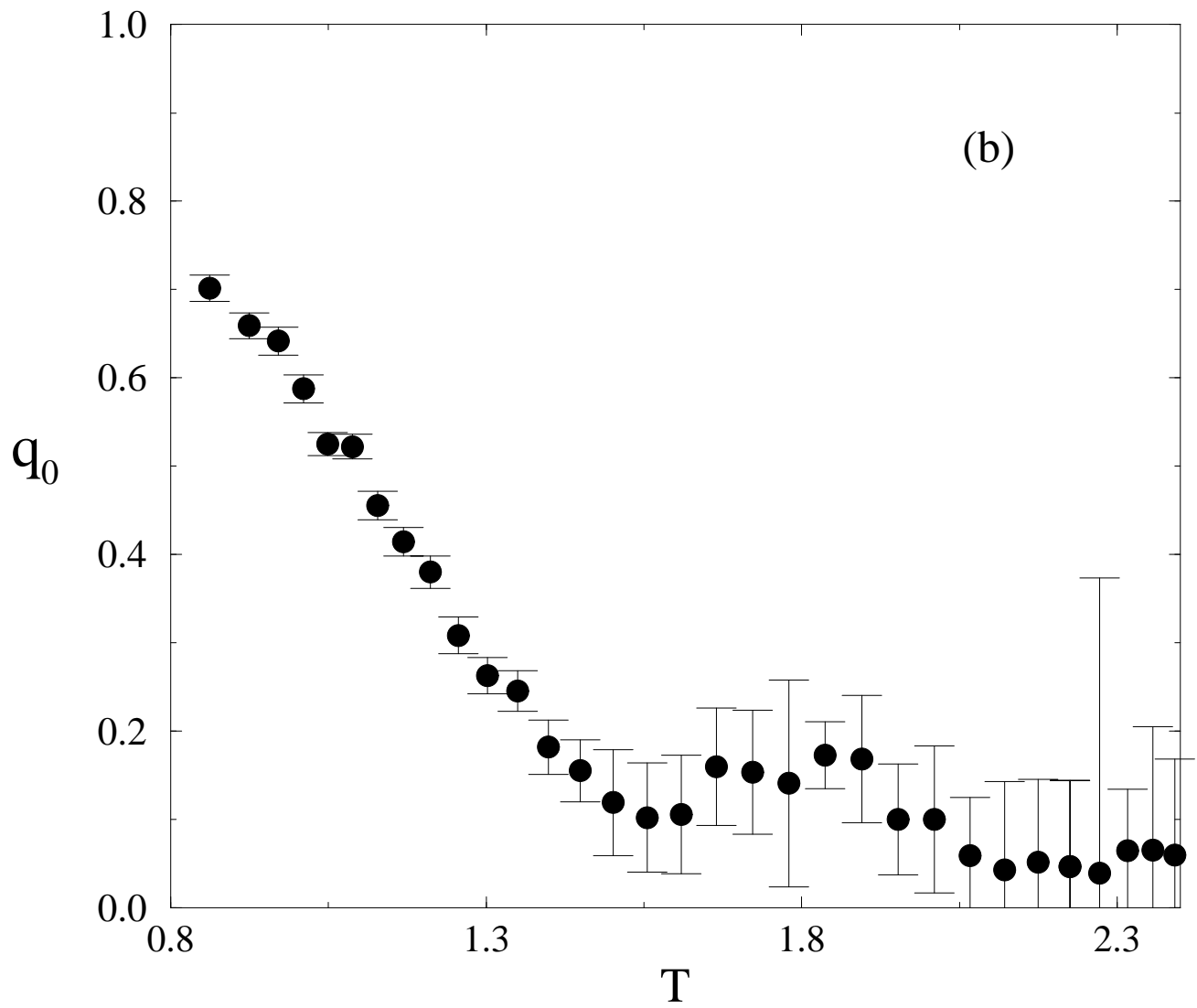
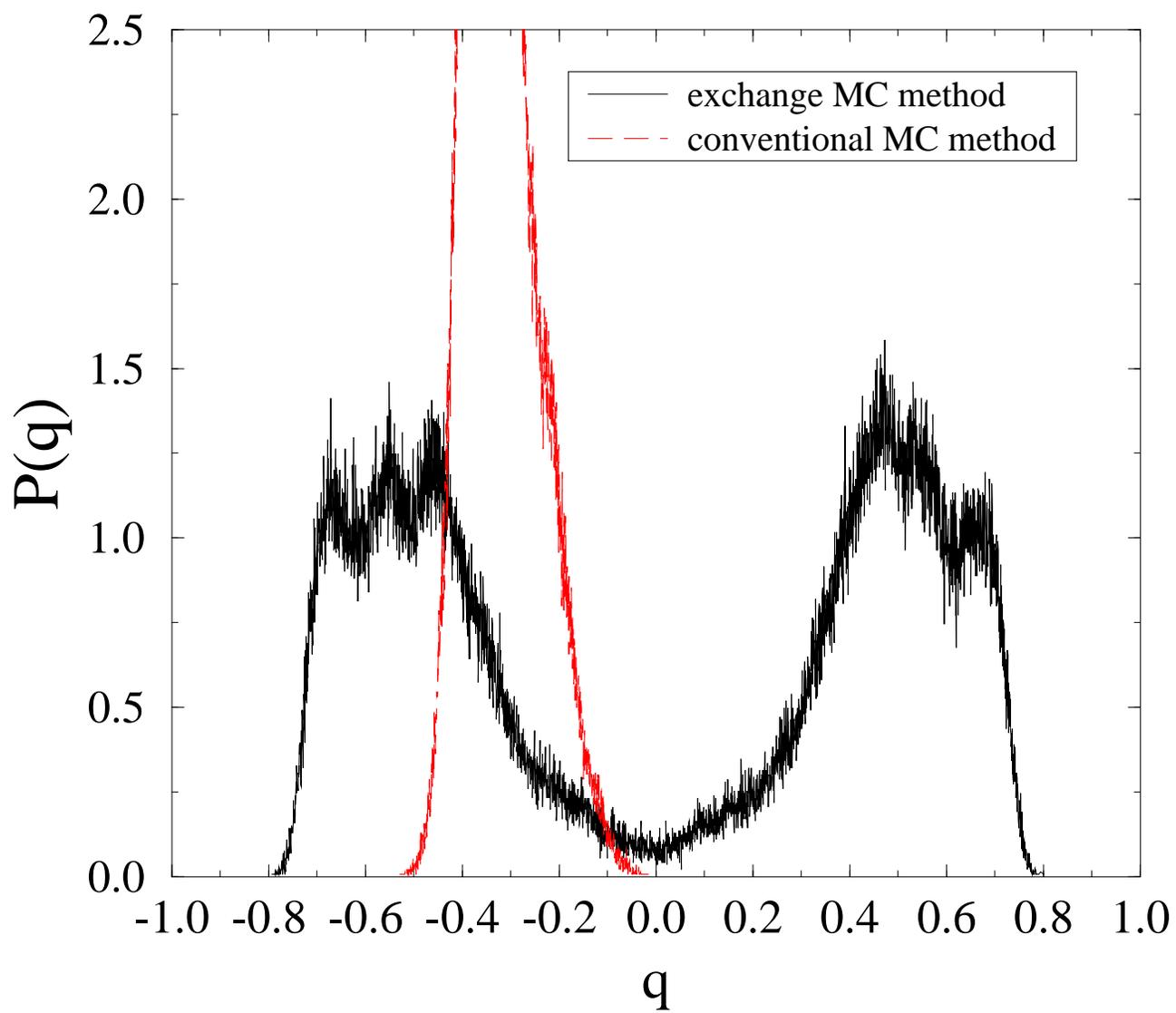

In this subsection we show the obtained distribution function $P_J(q)$
        of overlap function for a sample, which is defined by
\begin{equation}
        P_J(q) = \langle\delta(q-\frac{1}{N}\sum_{i}\sigma_i\tau_i)\rangle.
\end{equation}
The distribution function $P_J(q)$ is important to SG study because
        all physical quantities of our interest can be obtained from
        this function.
The distribution is expected to be an even function from the
        invariance of the Hamiltonian (\ref{eqn:model}) under the
        transformation $\sigma_i\rightarrow -\sigma_i$.
We demonstrate in Fig. \ref{fig:demo} that for $L=16$
        the distribution $P_J(q)$ is symmetric
        even below the SG transition temperature.
Note that only $3\times10^5$ MCS, which is about 10 times larger than
    $\tau_{\rm E}$
        are used to obtain Fig.~\ref{fig:demo}.
The distribution of the same sample obtained by the conventional MC method
        is also shown in Fig. \ref{fig:demo} with the broken line.
Obviously, it is far from symmetric
        and even the peak positions differ with each other.
It strongly suggests that
        the system does not reach the thermal equilibrium yet
        by the conventional MC method.

\section{discussion and summary}

One may consider that, for random systems,  $\{\beta_m\}$ should be
        determined for  each sample separately.
It is, however, not the case because the exchange probability $\{p_m\}$ is
        insensitive to bond realization as long as $\beta_{m+1}-\beta_{m}$ is
        sufficiently small.
It is contrasted to the case of the multi-canonical method, where the energy
        density should be estimated for each sample because it strongly depends
        on the structure of local minima in a sample. \cite{Berg92}

An advantage of the present method to the simulated tempering is that
        we need not estimate weighting factor (or free energy)
                to equidistribute the
        probability of visiting temperatures.
This is because the phase space of the present method
is a direct product of the original one
   in contrast with
        that of the simulated tempering, which is given by a direct sum.
Since the weight factor in the simulated tempering
 is an extensive parameter, the probability is very
        sensitive to it, and temperature traversing would be
        easily broken if the  estimation is not so good.

A more important merit of the present method
        appears in evaluating $P(q)$ of the SG system,
        where two copies of a system are independently simulated.
We have $M$ replicas for each copy, so that we can take $M(M-1)/2$
         un-correlated samplings of overlap in one simulation
         as compared with $M$ sampling is by
         $M$ independent runs of the simulated tempering.

As shown in the previous section,
the model to be simulated is not supposed to have any specific aspect,
so that we can, in principle, apply the present method to various models
        without modification.
Unfortunately the present method  seems not to work well
     in the systems exhibiting the first order phase transition.
Since the energy has a finite gap at transition point,
        the exchange between replicas below and above transition temperature
        hardly occurs in a large lattice, meaning that the condition (i)
        in section 3 is not satisfied.

In summary we have proposed a new MC method for simulating hardly
        relaxing systems.
We have applied this method to the three dimensional $\pm J$ Ising spin
        glass system, and
        found that the system really traverses over wide temperature space and
        the largest relaxation time in this dynamics is given
        by the ergodicity time,
        which is much smaller than the conventional one.
As a result, the order parameter distribution $P(q)$ can be  obtained
        below $T_{\rm c}$ up to $L=16$ within much shorter than the
        conventional MCS.
Numerical results of physical quantities and discussion on the  SG
        phase transions will be discussed in detail elsewhere.\cite{HK}
\acknowledgements
We are grateful to Mr. Y.  Iba for valuable discussion and comments
        on the extended ensemble method,
        and Professor H. Takayama for a critical reading of the manuscript
                and useful discussion.
Numerical calculations were mainly performed on Fujitsu VPP500
        at the Institute for Solid State Physics.
This work was supported by  Grand-in-Aid for Scientific Research on Priority
   Areas from the Ministry of Education, Science and Culture, Japan.
One of the authors (K.H) was supported by Fellowships of the Japan
  Society for the Promotion of Science for Japanese Junior Scientists.

\begin{figure}
\caption{Temperature dependence of the acceptance probability of
        exchange process.
The sample-to-sample errors are smaller than 1/10 of the size of
        symbols.}
\label{fig:accept}
\end{figure}

\begin{figure}
\caption{Ergodicity time $\tau_{\rm E}$
       as a function of the system size $V=L^3$.
The circles correspond to our data, and squares to multi-canonical method
by Berg et al. The unit of relaxation time is MC step per spins.}
\label{fig:tunnel}
\end{figure}
\begin{figure}
\caption{The modified autocorrelation function $q(t,\beta)$ for
        $1/\beta =T=0.862$, $1.088$, $1.256$, $1.399$, and $ 1.609$
( top to down).
}
\label{fig:qoft}
\end{figure}
\begin{figure}
\caption{ Temperature dependence of the relaxation time $\tau$ (a)
        and its amplitude $q_0$ (b) of the modified autocorrelation function
        (Fig. \protect{\ref{fig:qoft}}).
}
\label{fig:fit}
\end{figure}

\end{document}